\documentclass[10pt, final, conference, letterpaper, onecolumn, oneside]{IEEEtran}

\usepackage{cite}
\usepackage{graphicx}
\usepackage{subfigure}
\usepackage{amsmath}
\usepackage{amssymb}
\usepackage{epsfig}
\usepackage{times}
\usepackage{clrscode_mod}

\newtheorem{pavikc}{\em Corollary}
\newtheorem{pavikl}{\em Lemma}
\newtheorem{pavikt}{\em Theorem}

\newcommand{\argmin}{\operatornamewithlimits{argmin}}

\begin{document}

\title{Worst-Case Interactive Communication and Enhancing Sensor Network Lifetime}
\author{\IEEEauthorblockN{{\Large Samar Agnihotri and Pavan Nuggehalli}}\\
\IEEEauthorblockA{CEDT, Indian Institute of Science, Bangalore - 560012, India.\\
Email: \{samar, pavan\}@cedt.iisc.ernet.in}%
}

\maketitle

\begin{abstract}
We are concerned with the problem of maximizing the worst-case lifetime of a data-gathering wireless sensor network consisting of a set of sensor nodes directly communicating with a base-station. We propose to solve this problem by modeling sensor node and base-station communication as the interactive communication between multiple correlated informants (sensor nodes) and a recipient (base-station). We provide practical and scalable interactive communication protocols for data gathering in sensor networks and demonstrate their efficiency compared to traditional approaches.

In this paper, we first develop a formalism to address the problem of worst-case interactive communication between a set of multiple correlated informants and a recipient. We realize that there can be different objectives to achieve in such a communication scenario and compute the optimal number of messages and bits exchanged to realize these objectives. Then, we propose a formalism to adapt these results in the context of single-hop data-gathering sensor networks. Finally, based on this proposed formalism, we propose a clustering based communication protocol for large sensor networks and demonstrate its superiority over a traditional clustering protocol.
\end{abstract}

\section{Introduction}
\label{sec:Intro}
Many future and extant sensor networks feature tiny sensor nodes with modest energy resources, processing power, and communication abilities. A key networking challenge is to devise protocols and architectures that can provide relatively long operational sensor network lifetimes, in spite of these limitations. We define network lifetime as the time until the first sensor node or the base-station runs out of the energy. This reduces the network lifetime maximization problem to minimizing the maximum energy expenditure at sensor nodes and the base-station. Sensor nodes expend energy in sensing, computing, and communication. In this paper, we are mostly concerned with reducing the energy cost of communication. We neglect the energy consumed by the nodes in sensing and computing because sensing costs are independent of the communication strategy being deployed and computing costs are often negligible compared to communication costs.

The energy expended by a sensor node or the base-station in communication has two components: reception energy and transmission energy. The energy consumed in reception depends on the number of bits received and the per bit energy cost required to keep the receiver circuitry energized. The transmission energy depends on a number of factors such as transmit power levels, receiver sensitivity, channel state (including path loss due to distance and fading) and the kind of channel coding employed. In this paper, we assume that the data rates are low and that optimal channel coding is employed. Both these assumptions allow us to assume that the transmit power is linearly proportional to the data rate. Therefore, the communication energy is minimized by transmitting and receiving as few bits as possible.

In this paper, we first develop a theory of worst-case, lossless interactive communication between multiple correlated informants and a recipient. Then, assuming that the sensor data in a data-gathering sensor network is correlated, we model the communication between sensor nodes and the base-station in a single-hop data-gathering wireless sensor network as the $m$-message interactive communication between multiple correlated informants (sensor nodes) and a recipient (base-station), where at most $m$ messages are exchanged between a sensor node and the base-station. Interactive communication helps the sensor nodes in reducing their energy consumption by allowing those to use multiple compression rates while transmitting their information to exploit the correlation in sensor data and by offering computationally inexpensive encoding schemes. Based on our work on ``multiple correlated informants - single recipient'' interactive communication, we then propose a formalism to estimate the optimal number of messages and bits exchanged, in the worst-case, between the base-station and the sensor nodes in a data-gathering network. Then, we apply this formalism to maximize the worst-case lifetime of the network, for different objectives of communication. We conclude by proposing a new clustering protocol for large sensor networks, based on interactive communication.

To the best of our knowledge, our work for the first time addresses the problem of interactive communication between a recipient and a set of multiple correlated informants and then based on this formalism, proposes an alternative approach to enhance the lifetime of a data-gathering sensor network.

\section{Related Work}
\label{sec:relatedWork}
The ``multiple correlated informants - single recipient'' communication problem we are considering in this paper, is basically well-known distributed source coding (DSC) problem. This problem was first considered by Slepian and Wolf \cite{073slepianWolf} for lossless compression of discrete random variables and by Wyner and Ziv \cite{076wynerZiv} for lossy distributed compression. However, these work only provided theoretical bounds on the compression, but no method of constructing practical codes which achieve predicted theoretical bounds.

One of the essential characteristic of the standard DSC problem is that the information sources, also called encoders or informants, are not allowed to interact or cooperate with each other, for the purpose of compressing their information. There are two approaches to solve the DSC problem. First, allow the data-gathering node, also called decoder or recipient, and the informants to interact with each other. Second, do not allow the interaction between the recipient and informants. Starting with the seminal paper \cite{073slepianWolf}, almost all of the work in the area of DSC has followed the second approach. In the recent past, Pradhan and Ramchandran \cite{103pradhanRamchandran} and later \cite{101bajcsyMitran, 101graciaZhao, 102liverisXiong, 103graciaZhong, 104colemanLee, 104lanLiveris} have provided various practical schemes to achieve the optimal performance using this approach. An interested reader can refer to the survey in \cite{104xiongLiveris} for more information. However, only a little work \cite{092orlitsky, 104chouPetrovic}, has been done towards solving DSC problem when the recipient and the informants are allowed to interact with each other. Also, this work stops well short of addressing the general ``multiple correlated informants - single recipient'' interactive communication problem, which we are concerned with addressing in this paper.

In \cite{092orlitsky}, only the scenario in which two correlated informants communicate with a recipient is considered. It assumed that both the informants and recipient know the joint distribution of informants' data. Also, only the average of total number of bits exchanged is minimized. In \cite{104chouPetrovic}, only two messages are allowed to be exchanged between the encoder and a decoder, which may not be optimal for the general communication problem. Conversely, it does not address the problem of computing the optimal number of messages exchanged between the encoder and a decoder as well as the optimal number of bits sent by the encoder and a decoder for the given objective of the communication in an interactive communication scenario. Also, unlike \cite{092orlitsky}, this work concerns itself with the lossy compression at the encoders.

In our paper, in section \ref{sec:interactiveCommCompl}, we first provide various formulation of the worst-case, lossless ``multiple informants - single recipient'' interactive communication problem for various objectives of communication. Then, in section \ref{sec:worstcase}, we explicitly estimate corresponding optimal number of messages and bits transmitted by both, the recipient and informants. We assume that the joint probability distribution of informants' data is available only at the recipient. Previously, \cite{090orlitsky, 091orlitsky, 094zhangXia, 097ahlswedeCai} have addressed ``single informant - single recipient'' worst-case communication problem and attempted to bound its $m$-message complexity. In the same spirit, we attempt to solve here the worst-case ``multiple correlated informants - single recipient'' interactive communication problem.

In data-gathering sensor networks, the sensor data is assumed to be correlated and only the data-gathering node needs to learn about sensor data. This makes the data-gathering sensor networks a canonical problem to which DSC can be applied. A broad survey of the DSC without interaction schemes applied to sensor networks appear in \cite{104xiongLiveris}, and \cite{105zhongRabaey} makes a strong case for using \textit{asymmetric} DSC codes, such as Turbo code, LDPC codes, and convolutional codes in sensor networks. However, in the general sensor networks such proposals may not be optimal. For example, in the cluster-based sensor networks, where different sensor nodes alternatively assume the responsibility of data gathering, such proposals may be impractical, given the limited computational and energy resources of the nodes. In \cite{104chouPetrovic}, for the first time, it is proposed to use DSC with interaction in sensor networks to reduce the energy consumption at the sensor nodes. However, as mentioned earlier, their model of interactive communication is quite limited. Also, it does not directly relate the energy savings at the sensor nodes with the increase in sensor network lifetime. So, after introducing our system model for the sensor network in \ref{sec:systemModel}, in section \ref{sec:sensorNetLifetime}, we apply the ``multiple correlated informants - single recipient'' interactive communication formalism developed in the previous sections to maximize the worst-case operational lifetime of the data-gathering sensor networks. We conclude by providing a new clustering protocol, based on interactive communication, for the large sensor networks and the simulations results clearly establishing the efficiency of our approach.

A preliminary version of our ideas appears in \cite{isit07}, where we also extend the notions of \textit{ambiguity set} and \textit{ambiguity}, proposed in \cite{090orlitsky} and derive some of their properties. We intend to address the average-case communication problem and some other variations of the problem considered here, in future.

\section{Notation}
\label{sec:notation}
In this section, we introduce the notation that will be frequently used in the rest of this paper.
\begin{description}
\item [$S$:] the set of $N$ informants.
\item [$\cal X$:] finite, discrete alphabet set. $|{\cal X}| = n$.
\item [${\cal P}$:] $N$-dimensional discrete probability distribution, ${\cal P} = p(x_1, \ldots, x_N), x_i \in {\cal X}$.
\item [$\Pi$:] the set of all $N!$ schedules to poll $N$ informants.
\item [$\pi(i)$:] the informant that is polled at $i^{\textrm{th}}$ position in the schedule $\pi$.
\item [$A_{\pi(i)}$:] the set $\{\pi(1), \ldots, \pi(i-1)\}$ of informants who have already communicated their data to the recipient before the $i^{\textrm{th}}$ informant in the schedule $\pi$.
\item [$S_{X_{\pi(i)}|X_{\pi(1)}, \ldots, X_{\pi(i - 1)}}(x_{\pi(1)}, \ldots, x_{\pi(i - 1)})$:] \hspace{2.01in} the \textit{conditional ambiguity set} of the recipient in informant $\pi(i)$'s data, when the data vector $(X_{\pi(1)}, \ldots, X_{\pi(i - 1)}) = (x_{\pi(1)}, \ldots, x_{\pi(i - 1)})$. We denote it as $S_{X_{\pi(i)}|X_{A_{\pi(i)}}}$.
\item [$\mu_{X_{\pi(i)}|X_{\pi(1)}, \ldots, X_{\pi(i - 1)}}(x_{\pi(1)}, \ldots, x_{\pi(i - 1)})$:] \hspace{2.01in} the \textit{conditional ambiguity} $|S_{X_{\pi(i)}|X_{A_{\pi(i)}}}|$. We denote it as $\mu_{X_{\pi(i)}|X_{A_{\pi(i)}}}$.
\item [$\widehat{\mu}_{X_{\pi(i)}|X_{\pi(1)}, \ldots, X_{\pi(i - 1)}}$:] \hspace{0.9in} the \textit{maximum conditional ambiguity}. We denote it as $\widehat{\mu}_{X_{\pi(i)}|X_{A_{\pi(i)}}}$.
\end{description}

\section{``Multiple Informants - Single Recipient'' Communication Complexity}
\label{sec:interactiveCommCompl}
Let us consider two persons $P_{\cal X}$ and $P_{\cal Y}$ interactively communicating with each other. $P_{\cal X}$ observes the random variable $X_1 \in {\cal X}$ and $P_{\cal Y}$ observes a possibly dependent random variable $X_2 \in {\cal X}$\footnote{In general, $X_1 \in {\cal X}_1$ and $X_2 \in {\cal X}_2$, where ${\cal X}_1$ and ${\cal X}_1$ are discrete alphabet sets, with possibly different cardinalities. However, to keep the discussion simple, we assume henceforth that all the random variables take the values from same discrete alphabet ${\cal X}$, unless stated otherwise.}. Let us assume that only $P_{\cal Y}$ knows the joint distribution $p(x_1, x_2)$. In the worst-case, $P_{\cal Y}$ needs to send $\max(1, \lceil \log \log \widehat{\mu}_{X_1|X_2} \rceil)$ bits to $P_{\cal X}$  to help it send its information in $\lceil \log \widehat{\mu}_{X_1|X_2} \rceil$ bits to let $P_{\cal Y}$ learn about $X_1$. However, we soon show that in the optimal communication protocol the recipient $P_{\cal Y}$ needs to send many more bits than the number of bits given above. A \textit{meaningful} message from $P_{\cal X}$ to $P_{\cal Y}$ reduces the ambiguity about $P_{\cal X}$'s data at $P_{\cal Y}$.

In the following, we generalize this discussion to ``$N$ multiple correlated informants - single recipient'' communication problem and show that the interactive communication between the recipient and informants using \textit{prefix-free} messages and \textit{instantaneous decoding} \cite{wowmom05}, reduces this problem to a serial communication problem where the optimal schedule, in which $N$ ``single recipient - single informant'' communication problems are solved, is to be computed. We consider the communication problems with different objectives and provide optimal protocols for the worst-case communication for each of those problems.

Let us consider a set of $N$ multiple correlated informants \textit{interactively} communicating with a recipient, where the objective of communication is that the recipient must learn about each informant's data with no probability of error, but an informant may or may not learn about other informants' data.

Communication takes place over $N$ binary, error-free channels, where each channel connects an informant with the recipient. An informant and the recipient can interactively communicate over the channel connecting them by exchanging messages (finite sequences of bits determined by agreed upon, deterministic protocol), but the informants cannot communicate directly with each other (although, they can communicate indirectly via the recipient). So, if in an interactive communication protocol, the recipient and an informant exchange at most $m$-messages, then at most $Nm$ messages are exchanged before the recipient learns of all informants' data. Each bit communicated over any channel, in either direction is counted. We want to estimate the optimal number of messages and bits exchanged, in the worst-case, for such scenarios.

The problem of interactive communication between a single recipient and one or more informants has various variations, which are of interest depending on the communication scenario being modeled and some of these variations are already studied in existing literature. For example, depending on whether only the recipient knows the joint distribution of informants' data or both the recipient and informants know it, whether sum of the total number of bits communicated or the maximum number of bits communicated by any node is minimized, and whether this minimization is carried only over the set of informants or over the recipient and the set of informants, one can formulate different problems. There can be many more such variations, such as whether one considers lossless or lossy communication. However, in the present work we concern ourselves with some such variations.

In this work, we assume that the joint probability distribution ${\cal P} = p(x_1, \ldots, x_N), x_i \in {\cal X}$, of informants' data is \textit{only} known to the recipient. Contrast this with the communication scenarios considered in \cite{090orlitsky, 091orlitsky, 092orlitsky, 094zhangXia}, where both, the recipient and informant know the joint distribution. However, note that \cite{090orlitsky, 091orlitsky, 092orlitsky, 094zhangXia} only considered the ``single recipient - single informant'' communication problem. In the present work, we consider, the worst-case communication complexity of the four different problems corresponding to our assumption that only the recipient knows $\cal P$. It should be noted that this assumption can also be made in the communication scenarios where even the recipient does not know $\cal P$, as follows. Let us assume that at the beginning of the communication between the recipient and the informants, the recipient does not know $\cal P$. However, as the recipient would collect the information from the informants drawn from $\cal P$, it would eventually be able to estimate $\cal P$. For example, in \cite{104chouPetrovic} a linear predictive model is used to estimate the correlation structure. So, once the recipient has the estimate of $\cal P$, our formalism applies. We emphasize that we assume nothing about this distribution, except that it is a discrete distribution with finite alphabet. The underlying assumption of our work is that the correlation model is either already known to the recipient or can be learnt by it. However, in the communication scenarios where it is not so, our formalism does not apply.

Let $\widehat{R}_{\pi}$ denote the total number of bits transmitted by the recipient, under schedule $\pi$, to all the $N$ informants, in the worst case. Let $\widehat{I}_{\pi(i), R}$ denote the number of bits transmitted by the informant $\pi(i)$ to the recipient, in the worst case. Also, let $m$ denote the total number of messages exchanged between the recipient and an informant, before the recipient unambiguously learns of the informant's data. So, in the worst-case, we have the following four communication problems.
\begin{eqnarray}
&&\min_{m \ge 1} \min_{\pi \in \Pi} \max_{i = 1, \ldots, N}\widehat{I}_{\pi(i), R} \label{eqn:wfirstProb} \\
&&\min_{m \ge 1} \min_{\pi \in \Pi} \max(\widehat{R}_{\pi}, \max_{i = 1, \ldots, N}\widehat{I}_{\pi(i), R}) \label{eqn:wsecondProb} \\
&&\min_{m \ge 1} \min_{\pi \in \Pi} \sum_{i = 1}^{N}\widehat{I}_{\pi(i), R} \label{eqn:wthirdProb} \\
&&\min_{m \ge 1} \min_{\pi \in \Pi} (\widehat{R}_{\pi} + \sum_{i = 1}^{N}\widehat{I}_{\pi(i), R}) \label{eqn:wfourthProb}
\end{eqnarray}
Note that in above problem formulations, the first node $\pi(1)$ in any schedule $\pi$, sends its data uncompressed or at most compressed based on its past data. This node cannot exploit the data correlation structure to compress its data.

\section{Worst-Case Communication Complexity}
\label{sec:worstcase}
Let us consider a communication schedule $\pi \in \Pi$. Let us assume that the informants $\pi(1), \ldots, \pi(i-1)$ have already communicated their data to the recipient. Every informant $\pi(i)$ knows that it needs to send its data in at most $\lceil \log n \rceil$ bits to the recipient, where $n$ is the number of possible data values any informant's data can assume. The conditional ambiguity set of the recipient of informant $\pi(i)$'s data is $S_{X_{\pi(i)}|X_{A_{\pi(i)}}}$, with $\widehat{\mu}_{X_{\pi(i)}|X_{A_{\pi(i)}}} \le n$.

Before solving the problems in \eqref{eqn:wfirstProb}, \eqref{eqn:wsecondProb}, \eqref{eqn:wthirdProb}, and \eqref{eqn:wfourthProb}, we list without proof, the following properties of conditional ambiguity set, conditional ambiguity, and maximum conditional ambiguity, respectively.
\begin{eqnarray*}
S_{X_{\pi(i)}|X_{A_{\pi(i)}}}(x_{A_{\pi(i)}}) & = & \bigcap_{j = 1}^{i-1} S_{X_{\pi(i)}|X_{\pi(j)}}(x_{\pi(j)}) \\
\mu_{X_{\pi(i)}|X_{A_{\pi(i)}}}(x_{A_{\pi(i)}}) & \le & \min_{1 \le j \le i-1} \mu_{X_{\pi(i)}|X_{\pi(j)}}(x_{\pi(j)}) \\
\widehat{\mu}_{X_{\pi(i)}|X_{A_{\pi(i)}}} & \le & \min_{1 \le j \le i-1} \widehat{\mu}_{X_{\pi(i)}|X_{\pi(j)}}
\end{eqnarray*}

\subsection{Solution for \eqref{eqn:wfirstProb}}
\label{subsec:wfirstProb}
\textbf{\textit{Complexity of one-message communication:}} When the recipient and an informant are allowed to exchange only one message, then this message is from the informant to the recipient. As the informant in such scenario has no information about the ambiguity set of the recipient in its data, it sends $\lceil \log n \rceil$ bits to the recipient. In such situation, the solution to the problem in \eqref{eqn:wfirstProb} is trivial, as any order in which the informants communicate with the recipient results in an optimal communication schedule.

\textbf{\textit{Complexity of two-message communication:}} With the recipient and an informant allowed to exchange two messages, the recipient sends the first message to the informant, then based on its own information and the information contained in the recipient's message, the informant sends the second message to the recipient.

Given that the ambiguity set of the recipient of informant $\pi(i)$'s data is $S_{X_{\pi(i)}|X_{A_{\pi(i)}}}$, with maximum ambiguity $\widehat{\mu}_{X_{\pi(i)}|X_{A_{\pi(i)}}}$, in the worst-case, the recipient requires at least $\lceil \log \widehat{\mu}_{X_{\pi(i)}|X_{A_{\pi(i)}}} \rceil$ bits to learn unambiguously about $\pi(i)$'s data. So, it is both necessary and sufficient that $\pi(i)$ sends $\lceil \log \widehat{\mu}_{X_{\pi(i)}|X_{A_{\pi(i)}}} \rceil$ bits to the recipient. However, to help $\pi(i)$ send its information in just these many bits, the recipient informs it in $\widehat{\mu}_{X_{\pi(i)}|X_{A_{\pi(i)}}} \lceil \log n \rceil$ bits about those of its $n$ possible data values which belong to $S_{X_{\pi(i)}|X_{A_{\pi(i)}}}$. Then, $\pi(i)$ constructs the prefix-free codes corresponding to those data values and sends the code corresponding to its actual data value to the recipient in $\lceil \log \widehat{\mu}_{X_{\pi(i)}|X_{A_{\pi(i)}}} \rceil$ bits.

Following this protocol to poll all the informants, the total number of bits transmitted by recipient under schedule $\pi$, is
\begin{eqnarray}
\label{eqn:recipientWorstCase}
\widehat{R}_{\pi} \!\!\!\!\! & = & \!\!\!\!\! \sum_{i=1}^N \widehat{B}_{R, \pi(i)} \\
& = & \!\!\!\!\! \sum_{i=1}^N \widehat{\mu}_{X_{\pi(i)}|X_{A_{\pi(i)}}} \lceil \log n \rceil. \nonumber
\end{eqnarray}
The total number of bits transmitted by the informant $\pi(i)$ is
\begin{equation}
\label{eqn:informantsWorstCase}
\widehat{I}_{\pi(i), R} = \lceil \log \widehat{\mu}_{X_{\pi(i)}|X_{A_{\pi(i)}}} \rceil.
\end{equation}
$\widehat{R}_{\pi}$ bits are \textit{sufficient} for any model of correlation in the informants' data and \textit{necessary} too for some models of correlation.

\begin{pavikt}
\label{thrm:wfirstProb}
For $m \ge 2$, $\lceil \log \widehat{\mu}_{X_{\pi(i)}|X_{A_{\pi(i)}}} \rceil$ bits are both necessary and sufficient for the recipient to unambiguously learn about informant $\pi(i)$'s information.
\end{pavikt}
\begin{IEEEproof}
Omitted for brevity.
\end{IEEEproof}

\begin{pavikc}
\label{cor:wfirstProb}
Two messages are optimal.
\end{pavikc}
\begin{IEEEproof}
Previous theorem proves that $\lceil \log \widehat{\mu}_{X_{\pi(i)}|X_{A_{\pi(i)}}} \rceil$ bits from informant $\pi(i)$ are both necessary and sufficient for the recipient to learn about $\pi(i)$'s data. Also, each informant sends this optimum number of bits even when only two messages are allowed to be exchanged between the recipient and the informant $\pi(i)$. So, using the principle of Occam's razor, two messages are optimal.
\end{IEEEproof}

We are interested in finding the schedule $\pi^*$ that solves \eqref{eqn:wfirstProb}. However, Theorem \ref{thrm:wfirstProb} and Corollary \ref{cor:wfirstProb}, reduce it to
\begin{equation}
\label{eqn:worstCaseProblem}
\pi^* = \argmin_{\pi \in \Pi} \max_{i = 1, \ldots, N}\widehat{I}_{\pi(i), R}.
\end{equation}

The $\min\max$ nature of the problem in \eqref{eqn:worstCaseProblem} ensures that the \textit{Minimum Cost Next (MCN)} algorithm described below computes the optimal schedule in \eqref{eqn:worstCaseProblem}.

\hspace{-0.5cm}\hrulefill

\hspace{-0.25cm}{\textbf{Algorithm:} MCN}

\vspace{-0.2cm}\hspace{-0.5cm}\hrulefill
\begin{codebox}
\li Initialization: $k = 1$, $A_{\pi^{MCN}(k)} = \phi$.
\li \While $(k \leq N)$
\li $\pi^{MCN}(k) = \argmin_{i \in S-A_{\pi^{MCN}(k)}} \widehat{I}_{i, R}$. \label{li:next-node}
\li $A_{\pi^{MCN}(k+1)} = A_{\pi^{MCN}(k)} \cup \pi^{MCN}(k)$.
\li $k = k + 1$.
    \End
\end{codebox}
\vspace{-0.2cm}\hrulefill

\begin{pavikl}
\label{lemma:worstmcnMax}
\textit{MCN} schedule solves \eqref{eqn:worstCaseProblem}.
\end{pavikl}
\begin{IEEEproof}
We describe a procedure to modify a given schedule into another schedule such that value of the objective function does not increase. It will be apparent that iteratively applying this procedure on any schedule finally leads to the \textit{MCN} schedule $\pi^{MCN}$. Let $\pi^{OLD}$ be any schedule. Suppose it differs from $\pi^{MCN}$ first in the $m^{\textrm{th}}$ position, that is:
\begin{eqnarray}
\pi^{OLD}(k) & = & \pi^{MCN}(k), \quad 1 \le k \le m-1 \\
\pi^{OLD}(m) & \not= & \pi^{MCN}(m). \nonumber
\end{eqnarray}
Then there exists a number $l$ such that $\pi^{OLD}(l) = \pi^{MCN}(m), \quad l>m$. We construct a new schedule $\pi^{NEW}$ by modifying $\pi^{OLD}$ as follows:
\begin{eqnarray}
\label{newSchedule1}
\pi^{NEW}(k) & = & \pi^{MCN}(k), \quad 1 \le k \le m \\
\pi^{NEW}(k) & = & \pi^{OLD}(k-1), \quad m < k \le l \nonumber \\
\pi^{NEW}(k) & = & \pi^{OLD}(k), \quad l < k \le N \nonumber
\end{eqnarray}
In words, in $\pi^{NEW}$, we poll $\pi^{MCN}$ for first $m$-slots, followed by $\pi^{OLD}$ for next $N-m$ slots.

In order to establish that $\pi^{NEW}$ is at least as good as $\pi^{OLD}$, we need to show that
\begin{equation}
\label{weqn1}
\max_{i = 1, \ldots, N} \widehat{I}_{\pi^{NEW}(i), R} \le \max_{i = 1, \ldots, N} \widehat{I}_{\pi^{OLD}(i), R}.
\end{equation}
From \eqref{newSchedule1}, it follows that for $1 \le i \le m-1$ and $l+1 \le i \le N$
\begin{equation*}
\widehat{I}_{\pi^{NEW}(i), R} = \widehat{I}_{\pi^{OLD}(i), R}.
\end{equation*}
So, it suffices to show that
\begin{equation}
\label{weqn2}
\max_{i = m, \ldots, l} \widehat{I}_{\pi^{NEW}(i), R} \le \max_{i = m, \ldots, l} \widehat{I}_{\pi^{OLD}(i), R}.
\end{equation}
Using a lemma in \cite{isit07} that states that the conditioning reduces ambiguity, we have
\begin{equation}
\label{weqn3}
\max_{i = m+1, \ldots, l} \widehat{I}_{\pi^{NEW}(i), R} \le \max_{i = m+1, \ldots, l} \widehat{I}_{\pi^{OLD}(i), R}.
\end{equation}
Moreover, the \textit{MCN} construction ensures that
\begin{equation}
\label{weqn4}
\widehat{I}_{\pi^{NEW}(m), R} \le \widehat{I}_{\pi^{OLD}(m), R}.
\end{equation}
Equations \eqref{weqn3} and \eqref{weqn4}, imply \eqref{weqn2}, proving the lemma.
\end{IEEEproof}

\subsection{Solution for \eqref{eqn:wsecondProb}}
\label{subsec:wsecondProb}
\textbf{\textit{Complexity of one-message communication:}} The communication problem here is the same as the corresponding problem in subsection \ref{subsec:wfirstProb}. Every informant sends $\lceil \log n \rceil$ bits to the recipient and the recipient sends no bits and any order in which the informants communicate with the recipient results in an optimal communication schedule.

\textbf{\textit{Complexity of two-message communication:}} Using the two message protocol of previous subsection~\ref{subsec:wfirstProb}, we see that for every recipient-informant communication pair, the number of bits $\widehat{B}_{R, \pi(i)}$ transmitted by the recipient in communicating with informant $\pi(i)$ are always more than the number of bits $\widehat{I}_{\pi(i), R}$ transmitted by the informant $\pi(i)$. This implies that
\begin{equation*}
\widehat{R}_{\pi} > \max_{i = 1, \ldots, N}\widehat{I}_{\pi(i), R}.
\end{equation*}
So, in this case \eqref{eqn:wsecondProb} reduces to finding a schedule $\pi$ that minimizes $\widehat{R}_{\pi}$. However, as $\widehat{B}_{R, \pi(i)} > \lceil \log n \rceil$, so is $\widehat{R}_{\pi}$. So, the two-message complexity of this protocol for the problem in \eqref{eqn:wsecondProb}, is more than the one-message complexity. This implies that this two-message protocol is not optimal. In the following, we prove that there is no two-message protocol whose complexity is less than the complexity of the one-message protocol given above.

\begin{pavikt}
\label{thrm:wsecondProb}
There is no two-message protocol with complexity less than $\lceil \log n \rceil$.
\end{pavikt}
\begin{IEEEproof}
Omitted for brevity.
\end{IEEEproof}

\begin{pavikc}
\label{cor:wsecondProb}
One message protocol is optimal for the problem in \eqref{eqn:wsecondProb}.
\end{pavikc}
\begin{IEEEproof}
The proof follows from the last theorem.
\end{IEEEproof}

\subsection{Solutions for \eqref{eqn:wthirdProb} and \eqref{eqn:wfourthProb}}
\label{subsec:wthirdfourthProb}
Due to the paucity of the space, we do not discuss the optimal solutions for the problems in \eqref{eqn:wthirdProb} and \eqref{eqn:wfourthProb}.

\section{Sensor Network: System Model}
\label{sec:systemModel}
We consider a network of $N$ battery operated sensor nodes strewn in a coverage area. The nodes are assumed to interactively communicate with the base-station in a single hop. Sensor node $k, k \in \{1, \ldots, N\}$ has $E_k$ units of energy and the base-station has $E_{BS}$ units of energy. The wireless channel between sensor $k$ and the base-station is described by a symmetrical path loss $d_k$, which captures various channel effects and is assumed to be constant. This is reasonable for static networks and also for the scenarios where the path loss varies slowly and can be accurately tracked.

The network operates in a time-division multiple access (TDMA) mode. Time is divided into slots and in each slot, the base-station gathers data from every sensor node. Let us assume that the sensor data at every time slot is described by a random vector $(X_1, \ldots, X_N) \sim {\cal P}$. This distribution is \textit{only} known to the base-station. We assume the spatial correlation in the sensor data and ignore temporal correlation, as it can easily be incorporated in our work for data sources satisfying the Asymptotic Equipartition Property.

We assume static scheduling, that is the base-station uses the same sensor polling schedule in every time slot, until the network dies. The worst-case lifetime of a sensor node (base-station) under schedule $\pi \in \Pi$ is defined as the ratio of its total energy and its worst-case energy expenditure in a slot, under schedule $\pi$. However, as argued in Introduction, it is only the communication energy expenditure that we are here concerned with. We define network lifetime as the time until the first sensor node or the base-station runs out of the energy. This definition has the benefit of being simple, practical, and popular \cite{100changTassiulas} and as shown below, provides a neat and intuitive $\max\min$ formulation of the network lifetime in terms of the lifetimes of the sensor nodes and the base-station.

To model the transmit energy consumption at the base-station and the sensor nodes, we assume that transmission rate is linearly proportional to signal power. This assumption is motivated by Shannon's AWGN capacity formula which is approximately linear for low data rates. So, a node $k$ under schedule $\pi$ expends $B_{\pi(k)} d_k$ units of energy to transmit $B_{\pi(k)}$ units of information. Let $E_r$ denote the energy cost of receiving one bit of information. For simplicity, let us assume that it is same for both the base-station and the sensor nodes.

The general sensor network lifetime maximization problem is to solve joint source-channel coding problem for multi-access networks. However, we assume the separation between source and channel coding, though it is well-known that, in general, the source-channel separation does not hold for the multi-access joint source-channel coding problem \cite{CoverElGamal}. In this work, we assume that the optimal channel coding is employed. So, the general problem reduces to solving the distributed source coding problem to find the optimal rates (the number of bits to transmit), which maximize network lifetime. However, the optimal rate-allocation is constrained to lie within the Slepian-Wolf achievable rate region. This makes the problem computationally challenging. We simplify the problem by introducing the notion of \textit{instantaneous decoding} \cite{wowmom05} and thus reduce the optimal rate allocation problem to computing the optimal scheduling order, albeit at some loss of optimality. This loss of optimality occurs because, in general, turning a multiple-access channel into an array of orthogonal channels by using a suitable MAC protocol (TDMA in our case) is well-known to be a suboptimal strategy, in the sense that the set of rates that are achievable with orthogonal access is strictly contained in the Ahlswede-Liao capacity region \cite{Cover_book}.

\section{Maximizing Sensor Network Lifetime}
\label{sec:sensorNetLifetime}
To begin with, let us assume that the interaction between the base-station and the sensor nodes is not allowed. Then, in the worst-case, every node sends $\lceil \log n \rceil$ bits to the base-station to convey its information. However, if every node knows ${\cal P}$ and the data of all other nodes, then it only needs to send the bits describing its data conditioned on the data of the nodes already polled \cite{104cristescuLozano}. In the real single-hop sensor networks, neither it is possible that every node knows about all other nodes' data, given the limited communication capabilities of the sensor nodes; nor it is desired that the sensor nodes perform such computationally intense processing, given their limited computational and energy capabilities.

However, if we allow the interaction between the base-station and sensor nodes, then the nodes can still send less than $\lceil \log n \rceil$ bits, yet avoid above issues. In fact, this is precisely the ``multiple correlated informants - single recipient'' communication problem of section \ref{sec:interactiveCommCompl}. Using the results derived there and identifying the recipient as the base-station and informants as the sensor nodes, in the following, we attempt to maximize the worst-case lifetime of the single-hop sensor networks, for the given model of energy consumption and spatial correlation in the sensor data.

The base-station and a sensor node interactively communicate by exchanging optimal number of messages for the different communication problems, given in section \ref{sec:interactiveCommCompl}. To estimate the worst-case lifetime of the sensor networks with the given objective of communication, we use the protocols in \ref{sec:worstcase} for the base-station and sensor nodes communication. One of the major results of our work on the worst-case ``multiple informants - single recipient'' interactive communication problem is that for the formulations of this problem in \eqref{eqn:wfirstProb}-\eqref{eqn:wfourthProb}, it is the recipient which carries the most of the burden of communication and computation. So, in the context of the sensor networks, this implies that the corresponding role is played-out by the base-station. This reduces the energy consumption at the sensor nodes, hence enhancing their lifetimes, with concomitant increase in the network lifetime. For example, this is reasonable in the scenarios where the base-station is computationally and energy-wise more capable than the sensor nodes, as discussed in \cite{105zhongRabaey}. Still, it may not be infinitely more capable. So, in the network lifetime estimation problem, we consider the total communication (transmission and reception) energy expenditure at every sensor node as well as the base-station, to also include the situations where it is the base-station that runs out of the energy first.

\subsection{Worst-Case Network Lifetime}
\label{subsec:worstNetLife}
Let $\widehat{E}_{BS, \pi(i)}$ denote the energy that the base-station spends in communicating with node $\pi(i)$ in the worst-case, that is, it denotes the energy that the base-station spends in transmitting and receiving the bits from node $\pi(i)$, in the worst-case. So,
\begin{equation}
\widehat{E}_{BS, \pi(i)} = \widehat{B}_{BS, \pi(i)} d_i + \widehat{I}_{\pi(i), BS} E_r.
\end{equation}
Similarly, let $\widehat{E}_{\pi(i), BS}$ denote the energy that the node $\pi(i)$ spends in communicating with the base-station. So,
\begin{equation}
\widehat{E}_{\pi(i), BS} = \widehat{I}_{\pi(i), BS} d_i + \widehat{B}_{BS, \pi(i)} E_r.
\end{equation}
On substituting for $\widehat{B}_{BS, \pi(i)}$ and $\widehat{I}_{\pi(i), BS}$ from \eqref{eqn:recipientWorstCase} and \eqref{eqn:informantsWorstCase}, respectively, we have
\begin{eqnarray}
\widehat{E}_{BS, \pi(i)} \! - \widehat{E}_{\pi(i), BS} \!\!\!\!\! & = & \!\!\!\!\! \big(\widehat{\mu}_{X_{\pi(i)}|X_{A_{\pi(i)}}} \lceil \log n \rceil \label{eqn:energyDifference} \\
& & + \lceil \log \log \widehat{\mu}_{X_{\pi(i)}|X_{A_{\pi(i)}}} \rceil \nonumber \\
& & - \lceil \log \widehat{\mu}_{X_{\pi(i)}|X_{A_{\pi(i)}}} \rceil\big) (d_i - E_r). \nonumber
\end{eqnarray}
Assuming $d_i \ge E_r$, this implies that $\widehat{E}_{BS, \pi(i)} - \widehat{E}_{\pi(i), BS} \ge 0$, that is, the base-station spends more energy in communicating with node $\pi(i)$ than vice versa.

Given our definitions of the sensor node, the base-station, and the network lifetimes, the worst-case lifetime $\widehat{L}$ of the network is the solution to the following optimization problem
\begin{eqnarray}
\widehat{L} = \max_{\pi \in \Pi} \, \min \Big(\frac{E_{BS}}{\sum_{i=1}^N \widehat{E}_{BS, \pi(i)}} \,, \min_{i = 1, \ldots, N} \frac{E_{\pi(i)}}{\widehat{E}_{\pi(i), BS}}\Big) \label{eqn:worstLifetime1}, \\
\widehat{L}^{-1} = \min_{\pi \in \Pi} \, \max \Big(\frac{\sum_{i=1}^N \widehat{E}_{BS, \pi(i)}}{E_{BS}} \,, \max_{i = 1, \ldots, N} \frac{\widehat{E}_{\pi(i), BS}}{E_{\pi(i)}}\Big). \label{eqn:worstLifetime2}
\end{eqnarray}

As it was proven in section \ref{sec:worstcase} that the interaction helps in solving the problems \eqref{eqn:wfirstProb} and \eqref{eqn:wthirdProb}, in the following we estimate the network lifetime when the corresponding communication protocols are used in the network for the data-gathering. More precisely, we use the optimum two message communication protocol for the problem \eqref{eqn:wfirstProb}. However, before we discuss the general solution, let us consider its two special cases.

\textit{Case 1:} Let $E_{BS} = E_1 = \ldots = E_N = E$. This is so when $N+1$ identical sensors form a sensor cluster and one of those sensor nodes, is also chosen as the clusterhead. Then, the problem in \eqref{eqn:worstLifetime2} reduces to
\begin{equation*}
\widehat{L}^{-1} = \frac{1}{E}\min_{\pi \in \Pi} \, \max \Big(\sum_{i=1}^N \widehat{E}_{BS, \pi(i)} \,, \max_{i = 1, \ldots, N} \widehat{E}_{\pi(i), BS}\Big).
\end{equation*}
However, from \eqref{eqn:energyDifference}, we know that $\sum_{i=1}^N \widehat{E}_{BS, \pi(i)} \ge \max_{i = 1, \ldots, N} \widehat{E}_{\pi(i), BS}$, so above equation reduces to
\begin{equation}
\label{eqn:case1}
\widehat{L}^{-1} = \frac{1}{E}\min_{\pi \in \Pi} \sum_{i=1}^N \widehat{E}_{BS, \pi(i)}.
\end{equation}
In lemma \ref{lemma:mcnSum} below, we prove that the \textit{Minimum Cost Next} or \textit{MCN} algorithm described in \ref{sec:worstcase} computes the optimal lifetime for the optimization problem in \eqref{eqn:case1}.

\begin{pavikl}
\label{lemma:mcnSum}
\textit{MCN} schedule solves
\begin{equation}
\label{eqn:sumMCN}
\pi_{sum} = \argmin_{\pi \in \Pi} \frac{\sum_{i=1}^N \widehat{E}_{BS, \pi(i)}}{E_{BS}}
\end{equation}
\end{pavikl}
\begin{IEEEproof}
Changing the line~\ref{li:next-node} of the \textit{MCN} algorithm in \ref{sec:worstcase} to $\pi^{MCN}(k)=\argmin_{i \in S-A} \sum_{j \in A \cup i} \widehat{E}_{BS, j}$, we obtain a version of the \textit{MCN} algorithm that solves \eqref{eqn:sumMCN}. Then proof is identical to the proof of Lemma \ref{lemma:worstmcnMax} if in equations \eqref{weqn1}-\eqref{weqn4}, we make use of the following mappings:
\begin{eqnarray*}
\max_{i = 1, \ldots, N} & \longmapsto & \sum_{i = 1}^N, \\
\widehat{I}_{\pi^{NEW}(i), R} & \longmapsto & \widehat{E}_{BS, \pi^{NEW}(i)}, \\
\widehat{I}_{\pi^{OLD}(i), R} & \longmapsto & \widehat{E}_{BS, \pi^{OLD}(i)}.
\end{eqnarray*}
\end{IEEEproof}

\textit{Case 2:} Let $E_1 = \ldots = E_N = E$, but $E_{BS} \gg E$. This is so when the base-station is \textit{infinitely} more capable than any of the identical sensor nodes. Then, \eqref{eqn:worstLifetime2} reduces to
\begin{eqnarray}
\widehat{L}^{-1} \!\!\!\!\! & = & \!\!\!\!\! \frac{1}{E_{BS}}\min_{\pi \in \Pi} \max \! \Big(\sum_{i=1}^N \widehat{E}_{BS, \pi(i)}, \frac{E_{BS}}{E} \max_{i = 1, \ldots, N} \widehat{E}_{\pi(i), BS} \Big) \nonumber \\
 & = & \frac{1}{E}\min_{\pi \in \Pi} \max_{i = 1, \ldots, N} \widehat{E}_{\pi(i), BS}, \mbox{ for } E_{BS} \gg E. \label{eqn:case2}
\end{eqnarray}
In lemma \ref{lemma:mcnMax}, we prove that the \textit{Minimum Cost Next} algorithm above computes the optimal lifetime for the optimization problem in \eqref{eqn:case2} too.

\begin{pavikl}
\label{lemma:mcnMax}
\textit{MCN} schedule solves
\begin{equation}
\label{eqn:maxMCN}
\pi_{max} = \argmin_{\pi \in \Pi} \max_{i = 1, \ldots, N} \frac{\widehat{E}_{\pi(i), BS}}{E_{\pi(i)}}
\end{equation}
\end{pavikl}
\begin{IEEEproof}
Changing the line~\ref{li:next-node} of the \textit{MCN} algorithm in \ref{sec:worstcase} to $\pi^{MCN}(k)=\argmin_{i \in S-A} \frac{\widehat{E}_{i, BS}}{E_{i}}$, we obtain a version of the \textit{MCN} algorithm that solves \eqref{eqn:maxMCN}. Then proof is identical to the proof of Lemma \ref{lemma:worstmcnMax} if in equations \eqref{weqn1}-\eqref{weqn4}, we make use of the following mappings:
\begin{eqnarray*}
\widehat{I}_{\pi^{NEW}(i), R} & \longmapsto & \frac{\widehat{E}_{\pi^{NEW}(i), BS}}{E_{\pi^{NEW}(i)}}, \\
\widehat{I}_{\pi^{OLD}(i), R} & \longmapsto & \frac{\widehat{E}_{\pi^{OLD}(i), BS}}{E_{\pi^{OLD}(i)}}.
\end{eqnarray*}
\end{IEEEproof}

The general problem in \eqref{eqn:worstLifetime1} or equivalently in \eqref{eqn:worstLifetime2} can be solved as follows. It follows from Lemmas \ref{lemma:mcnSum} and \ref{lemma:mcnMax} that $\pi_{sum}$ and $\pi_{max}$ are the \textit{MCN} schedules which optimally solve \eqref{eqn:sumMCN} and \eqref{eqn:maxMCN}, respectively. Let $S^{MCN} = \{\pi_{sum}, \pi_{max}\}$. Then, \eqref{eqn:worstLifetime2} reduces to:
\begin{equation}
\label{eqn:optimalSoln}
\widehat{L}^{-1} = \min_{\pi \in S^{MCN}} \, \max \Big(\frac{\sum_{i=1}^N \widehat{E}_{BS, \pi(i)}}{E_{BS}} \,, \max_{i = 1, \ldots, N} \frac{\widehat{E}_{\pi(i), BS}}{E_{\pi(i)}}\Big).
\end{equation}

\begin{pavikt}
\label{thrm:optimalSoln}
$\widehat{L}^{-1}$ in \eqref{eqn:optimalSoln} is optimal.
\end{pavikt}
\begin{IEEEproof}
We prove the theorem by contradiction. Let $\pi^* \not \in \{\pi_{sum}, \pi_{max}\}$ be the optimal schedule. Without the loss of any generality, let us assume that the schedule $\pi_{max}$ minimizes the RHS of \eqref{eqn:optimalSoln}. However, with $\pi^*$ being the optimal schedule, we have
\begin{equation}
\label{thrm:worsteqn1}
\max \Big(\frac{\sum_{i=1}^N \widehat{E}_{BS, \pi^*(i)}}{E_{BS}} \,, \max_{i = 1, \ldots, N} \frac{\widehat{E}_{\pi^*(i), BS}}{E_{\pi^*(i)}}\Big) \le \max \Big(\frac{\sum_{i=1}^N \widehat{E}_{BS, \pi_{max}(i)}}{E_{BS}} \,, \max_{i = 1, \ldots, N} \frac{\widehat{E}_{\pi_{max}(i), BS}}{E_{\pi_{max}(i)}}\Big).
\end{equation}

For the sake of simplicity, let us use the following notation:
\begin{eqnarray*}
\Sigma_{\pi^*} & = & \frac{\sum_{i=1}^N \widehat{E}_{BS, \pi^*(i)}}{E_{BS}}, \\
\Sigma_{\pi_{max}} & = & \frac{\sum_{i=1}^N \widehat{E}_{BS, \pi_{max}(i)}}{E_{BS}}, \\
\mbox{Max}_{\pi^*} & = & \max_{i = 1, \ldots, N} \frac{\widehat{E}_{\pi^*(i), BS}}{E_{\pi^*(i)}}, \\
\mbox{Max}_{\pi_{max}} & = & \max_{i = 1, \ldots, N} \frac{\widehat{E}_{\pi_{max}(i), BS}}{E_{\pi_{max}(i)}}.
\end{eqnarray*}

Once more without the loss of any generality, let us assume that for the schedule $\pi_{max}$
\begin{equation*}
\Sigma_{\pi_{max}} < \mbox{Max}_{\pi_{max}}.
\end{equation*}
This along with \eqref{thrm:worsteqn1} implies the six possibilities of relative ordering among $\Sigma_{\pi^*}, \mbox{Max}_{\pi^*}, \Sigma_{\pi_{max}}, \mbox{Max}_{\pi_{max}}$, as in figure \ref{fig1}.

\begin{figure}[t!]
\begin{center}
\includegraphics[width=8.85cm]{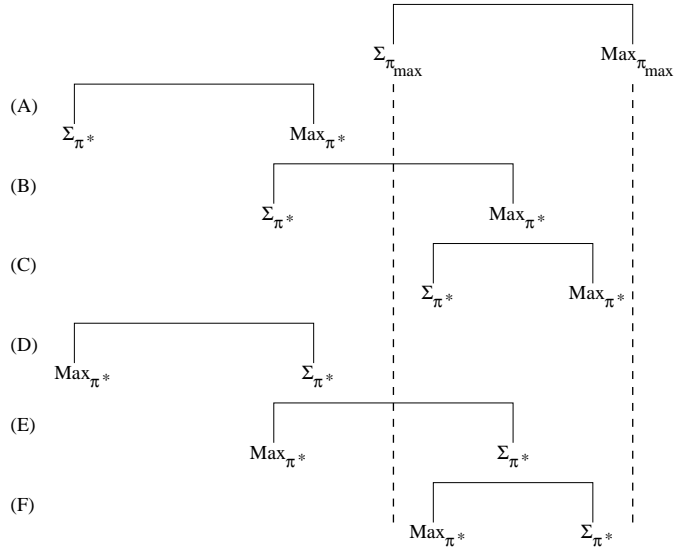}
\end{center}
\caption{Various orderings possible among $\Sigma_{\pi^*}$, $\mbox{Max}_{\pi^*}$, $\Sigma_{\pi_{max}}$,  $\mbox{Max}_{\pi_{max}}$, with $\Sigma_{\pi_{max}} <  \mbox{Max}_{\pi_{max}}$.}
\label{fig1}
\end{figure}

It is obvious from figure \ref{fig1} that in all cases, $\mbox{Max}_{\pi^*} <  \mbox{Max}_{\pi_{max}}$. So, using the sequence of steps in Lemma \ref{lemma:mcnMax}, we can iteratively convert the schedule $\pi^*$ to the schedule $\pi_{max}$, without any loss of its optimality, proving the optimality of $\pi_{max}$ with respect to maximizing the RHS of \eqref{eqn:optimalSoln}.
\end{IEEEproof}

\section{A New Communication Protocol for Sensor Networks Based on Interactive Communication}
\label{sec:interactiveProtocol}
In the section \ref{sec:interactiveCommCompl}, we mentioned that the assumption of \textit{instantaneous decoding} reduces the ``multiple correlated informants - single recipient'' communication problem to a serial communication problem, where the recipient only after retrieving the complete information from one informant, polls the next informant in the polling schedule. So, for $N$ informants, $N$ rounds of information gathering are serially executed. Using this protocol in the single-hop sensor networks introduces delay in data-collection at the base-station, which grows at least as $N$. This delay may be tolerable for small sensor networks, but most probably not for the large networks. In this section, we propose a low-delay communication protocol for arbitrarily large networks, based on the LEACH protocol \cite{100heinzelmanChandra}. Our protocol is same as LEACH in the cluster formation step, but differs from it in the data gathering step. So, in the proposed protocol, within a cluster, the clusterhead and sensors nodes communicate interactively using the formalism developed in the section \ref{sec:sensorNetLifetime}. As the data collection by the clusterheads in all clusters proceeds in parallel, this keeps the overall data-gathering delay at the base-station bounded.

In LEACH, the sensor nodes do not compress their data, so if each sensor node's data is derived from some finite set with cardinality $n$, then every sensor node sends $\lceil \log n \rceil$ bits to the clusterhead and the clusterhead compresses the data and sends it to the base-station. The achievable compression-ratio $r$ depends on the application and the type of data being sensed.

Our protocol, like LEACH, can be extended to form hierarchical clusters in very large sensor networks. In such networks, the clusterhead nodes interactively communicate with super-clusterhead nodes and so on until the top layer of the hierarchy, at which point the data is communicated to the base station. Then, this hierarchy can save a large amount of energy, yet keep the data-gathering delay within tolerable bounds.

\section{Simulation Results}
\label{sec:results}
For the purposes of modeling and performance simulations, we assume that the sensor network consists of $N$ sensor nodes uniformly distributed over a circle of radius $R$. The base-station is at the center of the circle. Each sensor node has at most $n$ bits of data to send to the base-station.

\subsection{Correlation model}
\label{subsec:correlModel}
As the model of the spatial correlation in sensor data, let us consider the first model of spatial correlation in sensor data introduced in \cite{correlModels}, with $\alpha_1 = 1.0, \beta_1 = 1.0$. So, let us define $B(X_i/X_j)$, the number of bits that the node $i$ has to send when the node $j$ has already sent its bits to the base-station, as follows:
\begin{equation}
\label{eqn:cor1}
B(X_i/X_j) = \left\{
                    \begin{array}{ll}
                     \lceil d_{ij} \rceil \mbox{ if } d_{ij} \le n\\
                     n \mbox{ if } d_{ij} > n,
                    \end{array}
             \right.
\end{equation}
where $X_i$ be the random variable representing the sampled sensor reading at node $i\in \{1, \ldots, N\}$, $n$ is the maximum number of bits that a node can send, and $d_{ij}$ denotes the distance between nodes $i$ and $j$.

Let us define $B(X_i/X_1, \ldots, X_{i-1})$, the conditional information when more than one node has already sent its information to the base-station, as follows:
\begin{equation}
\label{eqn:gc_ss4}
B(X_i/X_1, \ldots, X_{i-1}) = \min_{1 \le j < i} B(X_i/X_j)
\end{equation}

\subsection{Comparisons with LEACH}
\label{subsec:mcnVleach}
\begin{figure}[t!]
\begin{center}
\includegraphics[angle=-90, width=8.85cm]{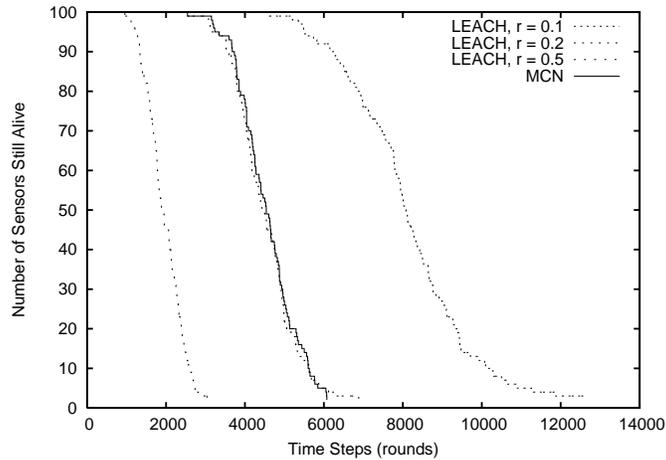}
\end{center}
\caption{Network lifetime comparison between MCN and LEACH with various compression ratios.}
\label{fig2}
\end{figure}

\begin{figure}[t!]
\begin{center}
\includegraphics[angle=-90, width=8.85cm]{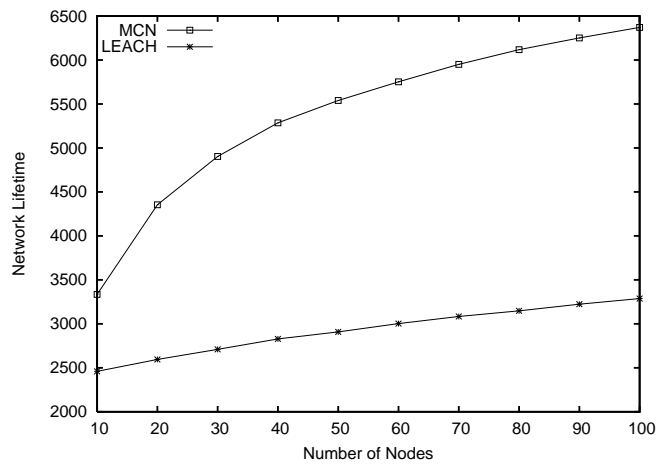}
\end{center}
\caption{Network lifetime comparison between MCN and LEACH protocols.}
\label{fig3}
\end{figure}

In this subsection, we compare the performance of the interactive communication protocol proposed in \ref{sec:interactiveProtocol} with the performance of LEACH protocol. Figures \ref{fig2} and \ref{fig3} show that our proposed protocol, denoted as ``MCN'', performs much better than LEACH for compression ratio $r > 0.2$.

Here we define the network lifetime to be number of data gathering rounds in which more than two nodes in the network are alive. In other words, the network is called dead when only two nodes are alive, one of these nodes in the clusterhead and other one is the sensor node.

Figure \ref{fig2} plots the number of sensor nodes which are still alive at the end of a certain number of data gathering rounds. In this plot, we compare the performance of our proposed protocol against LEACH protocol with compression-ratios, $r = 0.1, 0.2, 0.5$. The network started out with $N = 100$ nodes. This plot shows that as long as $r > 0.2$, the proposed protocol performs better than LEACH.

Figure \ref{fig3} compares the average achievable network lifetime for our proposed protocol and LEACH for different number of nodes in the network. For LEACH, we have set $r = 0.5$. Every data point in the plot corresponds to the network lifetime for the given number of nodes, averaged over $1000$ instances. Note that as the number of nodes in the network increases, the achievable lifetime increases accordingly, but saturates at some value. For our proposed protocol, the increase occurs due to a couple of reasons. Firstly, as the number of nodes increases in the given geographical area, the distance between the sensor nodes and the clusterhead those are associated with, decreases. Secondly, this increased node density also increases the correlation in the sensor data, so every node has to send fewer bits to the clusterhead. So, as the number of nodes increases, each sensor node transmits fewer bits over smaller distances, on average. However, as LEACH does not exploit the correlation in sensor data to reduce the transmission energy budget of the sensor nodes, the increase in the network lifetime with it comes only from the decreasing average distance of the sensor nodes from their respective clusterheads.

\section{Conclusions and Future Work}
\label{sec:conclusions}
In this work, we have considered ``multiple correlated informants - single recipient'' interactive communication problem, assuming that only the recipient knew of the correlation structure of the informants' data. However, if we assume that informants also know the correlation structure, then the optimal number of bits exchanged can be significantly reduced, resulting in more efficient communication protocols. Also, we have only presented the worst-case analysis in this paper. However, in some communication scenarios, it may be more desirable to estimate the optimal number of messages and bits exchanged, on average. We are presently working on such extensions of our work and their application to sensor networks.

\end{document}